\documentclass{vgtc}                          % final (conference style)
\ifpdf%                                % if we use pdflatex
  \pdfoutput=1\relax                   % create PDFs from pdfLaTeX
  \usepackage{graphicx}                % allow us to embed graphics files
  \DeclareGraphicsExtensions{.pdf,.png,.jpg,.jpeg} % for pdflatex we expect .pdf, .png, or .jpg files
\else%                                 % else we use pure latex
  \usepackage{graphicx}                % allow us to embed graphics files
  \DeclareGraphicsExtensions{.eps}     % for pure latex we expect eps files
\fi%

\graphicspath{{figures/}{pictures/}{images/}{./}} % where to search for the images

\usepackage{microtype}                 % use micro-typography (slightly more compact, better to read)
\PassOptionsToPackage{warn}{textcomp}  % to address font issues with \textrightarrow
\usepackage{textcomp}                  % use better special symbols
\usepackage{mathptmx}                  % use matching math font
\usepackage{times}                     % we use Times as the main font
\usepackage{cite}                      % needed to automatically sort the references
\usepackage{tabu}                      % only used for the table example
\usepackage{booktabs}                  % only used for the table example
\usepackage{soul}
\usepackage{todonotes}
\usepackage{balance}

\onlineid{0}

\vgtccategory{Research}

\vgtcinsertpkg

\title{Above Surface Interaction for Multiscale Navigation in \\Mobile Virtual Reality}

\author{Tim Menzner\thanks{e-mail: tim.menzner@stud.hs-coburg.de} %
\and Travis Gesslein\thanks{e-mail: travis.gesslein@hs-coburg.de} %
\and Alexander Otte\thanks{e-mail: alexander.otte@hs-coburg.de} %
\and Jens Grubert\thanks{e-mail: jens.grubert@hs-coburg.de}}
\affiliation{\scriptsize Coburg University of Applied Sciences and Arts}

\teaser{
	\centering
	\includegraphics[width=\columnwidth]{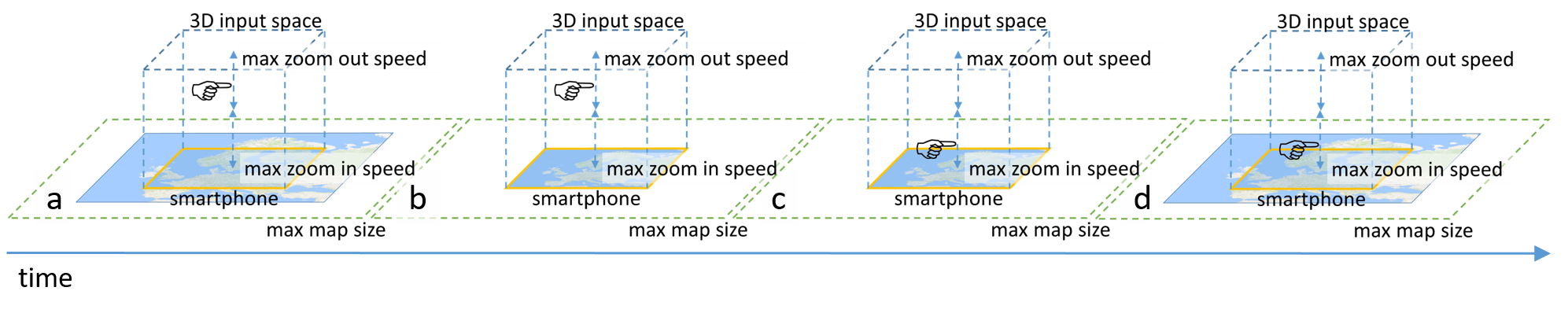}
	\caption{Relative, rate-controlled mapping for multiscale navigation between a touchscreen and VR headset: a) A user points her finger in the zoom out space (between the maximum finger height and halfway down the minimum height). b) While she holds her finger in this zoom out space the virtual information space scales down until it eventually reaches the size of the touch screen (with a rate corresponding to the height of the finger). c) The user points her finger in the zoom in space (between the minimum height and halfway up to the maximum height). c) While she holds her finger in this zoom in space the virtual information space scales up until it eventually reaches 1:1 scale (with a rate corresponding to the height of the finger). Target selection happens through touch on the touch screen.}
	\label{fig:relativemapping}
}

\abstract{Virtual Reality enables the exploration of large information spaces. In physically constrained spaces such as airplanes or buses, controller-based or mid-air interaction in mobile Virtual Reality can be challenging. Instead, the input space on and above touch-screen enabled devices such as smartphones or tablets could be employed for Virtual Reality interaction in those spaces. 

In this context, we compared an above surface interaction technique with traditional 2D on-surface input for navigating large planar information spaces such as maps in a controlled user study (n = 20). We find that our proposed above surface interaction technique results in significantly better performance and user preference compared to pinch-to-zoom and drag-to-pan when navigating planar information spaces.
} % end of abstract

\CCScatlist{
  \CCScatTwelve{Human-centered computing}{Visu\-al\-iza\-tion}{Visu\-al\-iza\-tion techniques}{Treemaps};
  \CCScatTwelve{Human-centered computing}{Visu\-al\-iza\-tion}{Visualization design and evaluation methods}{}
}

\begin{document}

%% The ``\maketitle'' command must be the first command after the
%% ``\begin{document}'' command. It prepares and prints the title block.

%% the only exception to this rule is the \firstsection command
%\firstsection{Introduction}

\maketitle

\section{Introduction} 

Recent progress in Virtual Reality (VR) technology has enabled the possibility of allowing users to conduct knowledge work in mobile scenarios. While the vision of a spatial user interface supporting knowledge work has been investigated for many years (e.g.~\cite{raskar1998office, rekimoto1999augmented}), the recent emergence of consumer-oriented VR headsets now make it feasible to explore the design of portable user interface solutions \cite{grubert2018office}. 

Lately, commercial VR head-mounted displays (HMDs) are progressing to 'inside-out' tracking using multiple built-in cameras. Inside-out tracking allows a simple setup of the VR system, and the ability to work in un-instrumented environments. Further, 3D finger tracking might be supported by traditional computing devices such as smartphones \cite{wang2016interacting}, which at the same time provide complementary high accuracy 2D touch input.

Given the capability of spatial hand and finger sensing, it is possible to see VR extending the input space of existing computing devices such as touch screens to a space around the devices, enhancing their usage. For example, knowledge workers can utilize existing 2D surface tools, new 3D tools, as well as using space around the devices and in front of the screen, reachable while sitting,  to represent and manipulate additional information. 

However, while VR is promising for mobile work it also brings its own additional challenges. For example, mobile information workers might many times be confined to a small space, such as the case of traveling in an airplane or a bus. While they can view potentially very large information spaces through their HMD, the ability to interact with those spaces might be limited in constrained physical spaces, restricting the suitability of ample spatial gestures \cite{grubert2018office}. Hence, we see it as important to examine interactions in potentially small input spaces. Also, while VR lends itself for interaction with three dimensional data, many knowledge worker tasks are still bound to two dimensional information surfaces \cite{grubert2018office}.

In this context, this paper explores how to efficiently navigate multiscale information spaces in a small input space using a combination of VR HMD and a touch surface.
Specifically, we explore the navigation of 2D planar information spaces such as maps using small spatial finger movements above a touchscreen while simultaneously allowing users to view beyond the boundaries of the touchscreen using a VR HMD. 

Our contribution lies in the design and evaluation of such as spatial interaction technique for navigation of large planar information spaces. In a user study (n = 20) we show that the proposed technique outperforms traditional 2D surface gestures.

\section{Related Work}
Our work builds on the areas of spatial and around device interaction as well as multiscale navigation, which will be reviewed next.

\subsection{Spatial Interaction}
%\subsection{Touch + Mid-Air and Touch + Gaze Interaction}

In the context of spatial user interfaces, a large number of techniques for selection, spatial manipulation, navigation and system control have been proposed \cite{laviola20173d}. Regarding object selection, Argelaguet et al. \cite{argelaguet2013survey} presented a survey on 3D selection techniques. For a recent survey on 3D virtual object manipulation we refer to Mendes et al. \cite{mendes2019survey}. Also, for spatial interaction across mobile devices (mostly in non-VR settings) recent surveys are available \cite{brudy2019cross, grubert2016challenges}.

For VR, spatial pointing techniques are of special interest, often relying on virtual hand or raycasting techniques. In the context of raycasting, Mayer et al. \cite{mayer2018effect} investigated the effects of offset correction and cursor on mid-air pointing and Schwind et al. \cite{schwind2018up} showed that avatar representations can have an effect on ray-casting-based pointing accuracy.  For virtual hand-based pointing, Barrera et al. \cite{barrera2019effect} indicated effects of stereo display deficiencies. Chan et al. \cite{chan2010touching} explored the use of visual and auditory feedback for facilitating spatial selection. Teather et al. \cite{teather2014visual} also investigated visual aids for 3D selection in VR. Lubos et al. \cite{lubos2014analysis} and Yu et al.\cite{yu2018target} investigated 3D selection in VR headsets, where Lubos found that visual perception has a larger effect on target acquisition than motor actions. Stuerzlinger and Teather proposed guidelines for targets in 3D pointing experiments \cite{stuerzlinger2014considerations}. The performance of 2D touch and mid-air selection on stereoscopic tabletop displays has been explored by Bruder et al. \cite{bruder2013touch, bruder2013touching}. Further, a number of techniques have been proposed for 3D object selection in the presence of object occlusion (e.g., \cite{feiner2003flexible, steed20043d, steinicke2006object, argelaguet2009efficient, bhowmick2018explorations}).

Besides, unimodal techniques, the combination of touch with mid-air, has drawn attention from researchers. For example, outside of VR, M{\"u}ller et al. \cite{muller2014mirrortouch} investigated the use of touch and mid-air interaction on public displays, Hilliges et al. \cite{hilliges2009interactions}  studied  tabletop settings. Multiple works have proposed to utilize handheld touchscreens in spatial user interfaces for tasks such as sketching, ideation and modelling (e.g., \cite{dorta2016hyve, ramanujan2016mobisweep, huo2017window, arora2018symbiosissketch, gasques2019pintar}), navigation of volumetric data \cite{song2011wysiwyf}, 3D data exploration \cite{lopez2015towards} or system control \cite{bowman2001design}. Spatial manipulation has seen particular interest in single-user settings (e.g., \cite{szalavari1997personal, mossel20133dtouch, marzo2014combining, babic2018pocket6, liang2013investigation, katzakis2015inspect, surale2019tabletinvr}) but also in collaborative settings \cite{grandi2017design}. 

Evolving from the magic lens \cite{bier1993toolglass} and tangible interaction concepts \cite{ullmer1997metadesk} tangible magic lenses allow to access and manipulate otherwise hidden data in interactive spatial environments. A wide variety of interaction concepts have been proposed within the scope of information visualization (for surveys we refer to \cite{tominski2014survey, tominski2017interactive}). Both rigid shapes (e.g., rectangular \cite{spindler2009paperlens}) or circular \cite{spindler2010tangible} and flexible shapes (e.g., \cite{steimle2013flexpad}) have been utilized as well as various display media (e.g., projection on cardboard \cite{spindler2009paperlens, chan2012magicpad}), transparent props \cite{schmalstieg1999using, brown2006magic}, handheld touchscreens \cite{grubert2015utility, leigh2015thaw}, or virtual lenses \cite{looser20073d} and sizes \cite{oh2006user}. Also, the combination of touch and gaze has seen recent interest for interaction in spatial user interfaces (e.g., \cite{pfeuffer2017gaze+, kyto2018pinpointing, schweigert2019eyepointing, ryu2019gg, hirzle2019design}).

Recently, Satriadi et al. and Austin et al. \cite{satriadi2019augmented, austin2020elicitation} investigated hand and foot gestures for map navigation within Augmented Reality in tabletop settings. Our work relates in the use of above surface interaction, but focuses on the interaction in constrained spatial environments.

Within this work, we concentrate on the use of mid-air interaction in conjunction with touch screen interaction for navigating potentially large planar information spaces such as maps.

\subsection{Around Device Interaction}
Along with the reduction of the size and weight of mobile and wearable devices, the need for complementary interaction methods evolved. 
%While the available space on a mobile device is continuously shrinking, research began investigating options for interaction
Research began investigating options for interaction
next to~\cite{Oakley:2014:IEO:2556288.2557138}, 
above~\cite{kratz2009hoverflow, Freeman:2014:TUA:2628363.2634215}, 
behind~\cite{DeLuca:2013:BAS:2470654.2481330, Wigdor:2007:LTS:1294211.1294259}, 
across~\cite{Schmidt:2012:CIS:2317956.2318005,chen2014duet}, or 
around~\cite{Zhao:2014:SDI:2642918.2647380,xiao2014toffee} 
the device. The additional modalities are either substituting or complementing the devices' capabilities. 

Different sensing solutions have been proposed such as infrared sensors (e.g., \cite{butler2008sidesight}), cameras (e.g., \cite{avrahami2011portico, yoo2016symmetrisense}, acoustic sensing (e.g., \cite{harrison2010appropriated, nandakumar2016fingerio, han2017soundcraft}, piezo sensors (e.g., \cite{xiao2014toffee}), depth sensors (e.g., \cite{kratz2012palmspace, sodhi2013bethere, chen2014air+}, electric field sensing (\cite{zhou2016aurasense}), or radar-based sensing (e.g., \cite{wang2016interacting}. Several approaches also investigated how to enable around-device interaction on unmodified devices (e.g., \cite{song2014air, song2015real, grubert2016glasshands, schneider2017towards}).

In VR, around device interaction has been studied in conjunction with physical keyboards for text entry (e.g., \cite{otte2019evaluating, otte2019towards, menzner2019capacitive, schneider2019reconviguration}).

This prior work, alongside commercialization efforts (e.g., for radar-based sensing \cite{wang2016interacting}, which recently made its way into commercial products such as Google Motion Sense\footnote{https://www.blog.google/products/pixel/new-features-pixel4/ Last access November 21st, 2019}) motivate us, that around-device interaction on mobile devices has the chance to become widely available for consumers.

\subsection{Multiscale Navigation}

To deal with large information spaces, multiscale interfaces \cite{Perlin:1993:PAA:166117.166125,  bederson1994pad++, guiard1999navigation} introduce the scale dimension, sometimes called Z\footnote{In this paper, we use the words scale and zoom interchangeably.}. Cockburn et al. \cite{cockburn2009review} give a comprehensive overview of overview+detail, zooming and focus+context interfaces including techniques for multiscale navigation. Many techniques make use of a separate zoom/scale dimension and a scroll/pan dimension to select a target value in the multiscale space \cite{guiard2004target}. Prior work indicated, that multiscale navigation with 2D interfaces obeys Fitt's law \cite{guiard1999navigation, guiard2004view}. Perluson et al. \cite{pelurson2016bimanual} investigated bimanual techniques for multiscale navigation. Appert et al. \cite{appert2006orthozoom} investigated 1D multiscale navigation with a mouse-based zooming technique.

In VR, multiscale collaborative virtual environments have been studied (e.g., \cite{zhang2005mcves, fleury2010generic, langbehn2016scale, le2016giant,  piumsomboon2018superman, piumsomboon2019shoulder}) and various metaphors (such as world-in-miniature) have been investigated for navigation tasks (e.g. \cite{kopper2006design, yan2016let, abtahi2019m}). Further, Zhang et al. \cite{stoakley1995virtual, laviola2001hands, zhang2008multiscale} introduced a progressive mental model supporting multiscale navigation and also indicated that navigation in multiscale virtual environments can be challenging \cite{zhang2009multiscale}.

For mobile devices, Jones et al. \cite{jones2012around} investigated mid-air interactions for multiscale navigation afforded by mobile depth sensors and discussed design implication for mid-air interactions and depth-sensor design. Kister et al. \cite{kister2017grasp} combined spatially-aware mobile displays and large wall displays for graph explorations. Kratz et al. \cite{kratz2010semi} proposed a semi-automatic zooming technique as an alternative to multi-touch zooming but could not indicate significant performance benefits. For mouse-based interaction in desktop environments, Igarashi and Hinckley proposed rate-based scrolling with automatic zooming to obtain a constant perceived scrolling speed in screen space. To this end, when the user scrolled fast, the view autonatically zoomed out. The authors indicated a comparable performance to the use of scroll bars.

Spindler et al. \cite{spindler2014pinch} compared pinch to zoom and drag to pan with mid-air spatial navigation utilizing absolute mappings and found that navigating a large information space by moving the display devices outperforms navigation with 2D surface gestures. Pahud et al. \cite{pahud2013toward} compared 2d surface gestures with spatial navigation but derived at different findings (spatial navigation was slower than 2D navigation). Related, Hasan et al. \cite{hasan2015comparing} compared dynamic peephole pointing and direct off-screen pointing with on-screen 2D gestures for map based analytic tasks. They found that while the spatially aware techniques resulted in up to 30\% faster navigation times, 2D onscreen gestures allowed for more accurate retrieval of work space content.

Our study also compares to a 2D surface gesture baseline technique, but compare it to relative-rate controlled  above surface interaction technique, which is suitable for a wide variety of usage scenarios, including interaction in physically constrained environments such as trains or touchdown places.

\section{Above Surface Interaction for Multiscale Navigation in Mobile VR}

% \begin{figure*}[t]
% 	\centering
% 	\includegraphics[width=2\columnwidth]{images/absoluteMapping02.png}
% 	\caption{Absolute, position-controlled mapping (not utilized in the final user study): a) A user points her finger at the maximum finger height. The information space is mapped to the bounds of the smartphone. b) the user points her finger halfway between the maximum finger height. The information space is mapped halfway between the bounds of the smartphone and its virtual 1:1 scale. c) the user points her finger at the minimum height. The information space is extended to 1:1 scale.}
% 	\label{fig:absolutemapping}
% \end{figure*}

Within this research, we set out to support efficient navigation of large information spaces using a potential small interaction space (such as the space on and above a smartphone or a tablet) using a combination of spatial interaction and a touch surface. In this joint interaction space between touchscreens and VR headsets, the touchscreen can be utilized for fine grained 2D touch interactions, while the VR headsets allows to visualize content beyond the device boundary of the touchscreen. Within this scenario, spatial tracking of users' hands and fingers could be achieved either by camera-based inside-out tracking on the VR headset (which at the same time could spatially track the position of the smartphone through marker-based or model-based tracking), by smartphone-integrated sensing, or a combination of both.

%\hl{... intro why multiscale is important ...
%... intro why we believe above surface sensing will become feasible ...
%... intro why interaction in constrained physical environments is important to be investigated ...
%} \\

Inspired by the large amount of prior work, we followed an iterative approach with multiple design loops consisting of conceptualization, implementation and initial user tests (dogfood tests) \cite{unger2012project, drachen2018games}. For prototyping purposes we relied on an external outside-in tracking system (OptiTrack Prime 13) in conjunction with a HTC Vive Pro headset.

%In order to find an optimal method for above surface interaction, we implemented and evaluated several possible techniques before choosing which to test in this study. 

First, we explored an absolute, position-controlled mapping of finger height to the scale of the virtual information space. %(see Figure \ref{fig:absolutemapping}, a-c). 
With this technique, the height of the finger above the smartphone display until a previous empirically determined maximum height (5 cm above the display) was directly mapped to a scale level of the virtual display.  %If the finger was on the display, the information space (in our case a virtual map) was at its maximum scale level. If the finger was at 5 cm (or above) the virtual information space was mapped to the smartphone screen size (assuming a matching aspect ratio of the smartphone screen and the virtual information space). 
The scaling changed linearly between minimum and maximum finger height. While initial user tests indicated that this mapping felt natural, the accuracy of the technique is dependent on the the maximum scale of the information space. 

%For the maximum zoom out factor, we compared a value where the map at maximum zoom out fit and filled out the display of the virtual smartphone representation with several values below, where the map at maximum zoom out was still to large to fit inside the display. 

To overcome this limitation, we implemented a relative, rate-controlled mapping, in which  the scale level was constantly increased or decreased with a certain speed dependent of the height of the finger over the display (see Figure \ref{fig:relativemapping}, a-d). For this approach, the volume above the smartphone display was separated into two smaller spaces of the same size. The first space extended from  above the display surface to half a previous defined maximum height. The second area extended from this height to the defined maximum height. If the finger was in the lower volume, the scale level increased. If the finger was in the upper volume, the scale level decreased. The point between those areas at half maximum height, theoretically would result in a scale change of 0. We experimented with a "dead zone" in the center (i.e. expanding the volume in which no scale change would occur). Initial user tests, indicated that the scale change around the center was sufficiently low that no extended dead zone was necessary. We also experimented with a dead zone directly above the touch surface but came to similar conclusions.

%Just like for the absolute zoom method, zoom was represented by changing the map size. 

For zooming out, a \textit{zoom base speed} parameter determining the maximum reachable zoom speed 
%(\hl{added parameter explanation later in own paragraph}) 
was multiplied with the finger height above the display normalized between half maximum height and maximum height. The result was used to decrease the map size by this factor. Similarly, for zooming in, the height was normalized between half maximum height and the minimum finger height, see Figure \ref{fig:relativemapping}. If the user touched the screen however, the zooming process was stopped so she could still conduct touchscreen input without constantly zooming in or moving on the other axes. %(\hl{information about "dead zone" where zoom speed = 0 included in figure and a few paragraphs up "The point between those areas at half maximum height, theoretically would result in a zoom speed of 0" }

The movement on the two other axes in both techniques was handled equally. The position of the finger on the x axis or respectively the y axis relative to the display middle point was multiplied with a \textit{plane base speed} parameter %(\hl{added parameter explanation later in own paragraph})
and was used to determine the speed in this direction, resulting in a rate-controlled change of the x,y position. In order to make sure that the user could always see where exactly the fingertip was hovering, a small flat white disc with a 5 mm diameter indicated where a ray cast perpendicular to the touch surface would hit the display. This point also functioned as the pivot point for zooming. 

Initial user tests indicated that the rate-controlled mapping resulted in both in better performance (i.e. task completion time) and was more preferred than the absolute one. Especially, on the larger map we tested (1.392m * 0.655m and a touchscreen width to virtual map width ratio of 1:13,2 compared to 13.92m * 6.55m and a ratio of 1:132), the user would have to zoom more to get to the minimum scale level. Specifically, the range between the display surface and a physically possible maximum finger height was too small to provide a smooth scale change. A small change in finger height would map to a large scale change. This made accurate and precise navigation very difficult. Compared to this, the relative movement was perceived much more smooth and precise on all map sizes. 

All base speed parameters were empirically determined. For the \textit{zoom base speed}, this value was 0.05 and for the \textit{plane base speed}, this value was 0.001. Please note that the zoom speed constant of 0.05 is dependent on the maximum finger height (in our case 5~cm) and the \textit{plane base speed} parameter is dependent on the smartphone display size. For other finger heights or display sizes those parameters can be linearly interpolated. Also, the maximum finger height of 5 centimeters above the display empirically produced the best results for various users.

%the Same is true for the maximum finger height, our tests showed that a maximum height of 5 centimeters above the display produced the best results.

\begin{figure*}[t!]
	\centering
	\includegraphics[width=2\columnwidth]{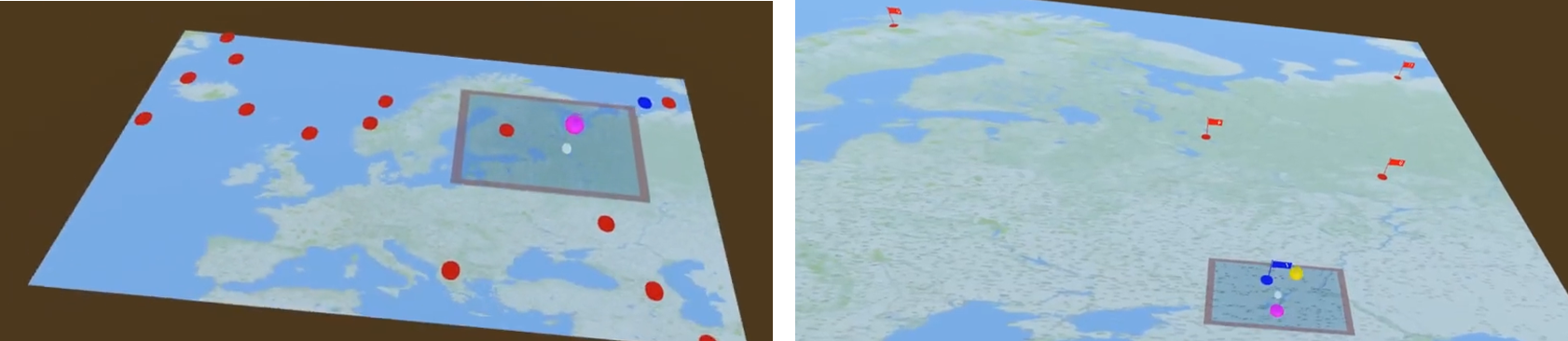}
	\caption{Left: View on the VR scene in the \textsc{3D Navigation} condition at a small scale. The active target is shown as blue dot, inactive targets as red dots. The pink sphere indicates the fingertip position, the white disc below the pivot point for zooming. The touch screen area of the smartphone is shown as semi-transparent blue rectangle with red border. Right: View on the VR scene in the \textsc{baseline} condition at 1:1 scale. The circular targets are complemented with 3D flag symbols to indicate that targets can be selected. The two fingertips used for zooming are indicated by yellow and pink spheres. The pivot point indicated by the white disc is visualized half-way between both fingertips. }
	\label{fig:conditions}
\end{figure*}

\section{User Study}

We aimed to compare the relative, rate-controlled mapping, which outperformed the absolute, position-controlled mapping in initial tests, with commonly used technique for navigating 2D information spaces on touch screens. To this end, we utilized the well known pinch-to-zoom and drag to pan interaction techniques and conducted a multiscale target search and selection task in a map navigation scenario.

\subsection{Study Design}
We followed a within-subjects design with two independent variables: \textsc{interface} and  \textsc{map size}. 

\textsc{interface} had two levels:  \textsc{3D Navigation} utilized the relative above surface interaction technique described in the previous section. The \textsc{baseline} condition utilized  2D pinch to zoom and drag to pan known from mobile map applications. The parameters for pan and zoom were chosen to mimic the Google Maps pan and zoom experience on the smartphone used in the user study (see Apparatus section), including inertia and pivot point for zooming. The \textsc{baseline} technique utilized the smartphone touch screen for sensing. We also verified that the navigation performance when wearing a VR HMD matches performance when not wearing an HMD but interacting with a mobile phone only version of the application.  A small white flat disc was used to visualize the pivot point of zooming, in this case the middle point between the two fingers used for zoom. The users' fingers used for interaction (typically thumb and index finger of the dominant hand) were visualized as colored spheres with 10 mm diameter (same as the target size).

There were two \textsc{map size}s: \textsc{small map} had a size of 1.45~m * 0.69~m resulting in ratio between touch screen width to virtual map width of 1:13.8. \textsc{large map} had a size of 144.71 * 69.11~m resulting in ratio between touch screen width to virtual map width of 1:1380. In a pilot study, we evaluated an additional map size (with a touchscreen width to virtual map width ratio of 1:138), but found no additional insights compared to the large map size and, hence, excluded it from the main study.

% and smartphone to virtual screen ratio of 1:1587331

%There were two \textsc{map size}s: \textsc{small map} had a size of 1.447m * 0.691m resulting in a smartphone to screen ratio of 1:158.71 \textsc{large map} had a size of 144.71 * 69.105 and smartphone to screen ratio of 1:1587331

%(ranging from \hl{XXX} too \hl{XXX} and found those two map sizes to be representative for \hl{... XXX}

Within each of the four conditions (2 \textsc{interface}s x 2 \textsc{map size}s, participants needed to select targets at three \textsc{target distance}s: \textsc{small distance}, \textsc{medium distance} and \textsc{large distance}. The order of \textsc{target distance} was randomized within each condition, ensuring that each of the three \textsc{target distances} appeared exactly five times per condition. The  target distances were defined relative to the length of the diagonal of each \textsc{map size} corresponding to 0.125 * map diagonal for \textsc{small distance}, 0.25 * map diagonal for \textsc{medium distance} and 0.5 * map diagonal for \textsc{large distance}. Hence, \textsc{small distance} corresponded to 0.2~m on the \textsc{small map} and 20~m on the \textsc{large map}. \textsc{medium distance} corresponded to 0.4~m on the \textsc{small map} and 40~m on the \textsc{large map}. \textsc{large distance} corresponded to 0.8~m on the \textsc{small map} and 80~m on the \textsc{large map}.

%\hl{more details on distances found in task. I am not sure where it fits better}

The conditions were blocked by \textsc{map size}, i.e. the starting order of \textsc{interface} was counter-balanced within each \textsc{map size} but both \textsc{interaction technique}s were conducted sequentially for a single \textsc{map size}. The order of \textsc{map size} was also counterbalanced (e.g., participant i starting with \textsc{map size} \textsc{small map}, participant i+1 starting with \textsc{large map}.

\begin{figure*}[ht]
	\centering
	\includegraphics[width=2\columnwidth]{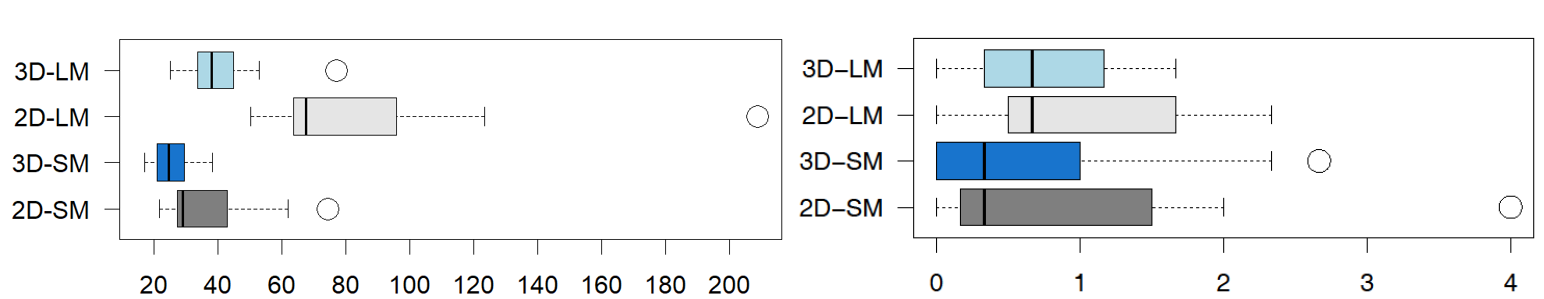}
	\caption{Left: Target acquisition time per target in seconds with significant differences between all conditions. Right: Number of errors with no significant differences between any condition. 2D: \textsc{baseline}, 3D: \textsc{3D Navigation}, SM: \textsc{small map}, LM: \textsc{large map}.The boxes encompass the interquartile range, the whiskers extend to 1.5 times the interquartile range, respectively the minimum and maximum data points, if those do not exceed the interquartile range. The solid vertical lines indicate the median.}
	\label{fig:performance}
\end{figure*}

\subsection{Task}

Participants were asked to conduct a target search and acquisition task. To this end, they had to identify a target on the map, navigate to the target and eventually select it. 

In each condition, the participants were asked to find and click a total of 15 circular targets with a diameter of 10 mm on a map. This target size was scale-independent (i.e. the target was always displayed with a 10 mm diameter), but the target could only be selected in the largest map scale (zoomed in to the maximum scale level). The active target was highlighted in blue, while all other potential targets were colored red. In order to click the target, the participant had to navigate close enough so that the target was inside the touch screen bounds (and at scale level 1:1). When this was achieved, it was possible to selecting it by dwelling the finger on the real world smartphone display at the target location for one second. This delay time was used to separate an accidental touch from an indented click. If the participant failed to hit the target at first attempt, an error was logged. In order to progress to the next marker, the participant still had to eventually achieve a successful hit afterwards. %Clicking a target was only possible when at maximum zoom in where the map and real world were mapped 1 to 1. 
%At minimum scale, the mapping was so that the whole map filled out the whole smartphone display. 
After successfully selecting the target, a new target was highlighted. The new target was always selected to fit a certain distance to the previous target, respectively the origin for the first target. Those distances were separated in three categories. Near, middle and far. Each target distance was randomly used  5 times adding up to 15 total targets per condition. The task started at the maximum scale level. 

%The exact distance in meters of each category was dependent on the map size. 
%The participants had to perform the same tasks on a total of two maps, each with the same aspect ratio as the smartphone but a different total area, namely 1 sqm for the small map compared to 10 000 sqm for the large map. 

%On both maps, the participants tested two different control options, the previous mentioned 3D Navigation Method and as a comparison 2D touch method with an 1 to 1 mapping, inspired by common controls in popular map navigation apps. 

\subsection{Procedure}
After welcoming, participants were asked to fill out a demographic questionnaire. This was followed by a short tutorial video, which gave first information about the to be fulfilled tasks and the two implemented interaction techniques. Thereafter, we conducted a training phase for the first condition, in which participants clicked 6 targets, 2 of each distance type, to get used to the controls. This training phase was later repeated before each condition with the respective technique. After finishing this training phase, the actual test condition was carried out.
After each condition, the participant filled out three questionnaires, the system usability scale SUS \cite{brooke1996sus}, the NASA Task Load Index TLX \cite{hart1988development} and the Simulator Sickness Questionaire SSQ \cite{kennedy1993simulator}. The conditions were blocked by \textsc{map size}. The starting order of the blocks (\textsc{small map} and \textsc{large map}) as well as the order of the \textsc{interface}s in each block were balanced across participants. After each block, the participants answered an additional questionnaire concerning their personal preferences related to both methods. After the first block, the participants had the possibility to take a 10 minute break before proceeding with the next block. Following both blocks and all questionnaires, the participants were asked to participate in a semi-structured interview, in which further feedback was collected. Finally, participants were thanked and compensated with a 10 Euro voucher.

The duration of each condition was strongly dependent on the speed of the participant, the map size and the used navigation method. Together with the questionnaires and the short interview, the overall average duration of the experiment was 60 minutes.

\subsection{Apparatus}

\begin{figure}[t!]
	\centering
	\includegraphics[width=\columnwidth]{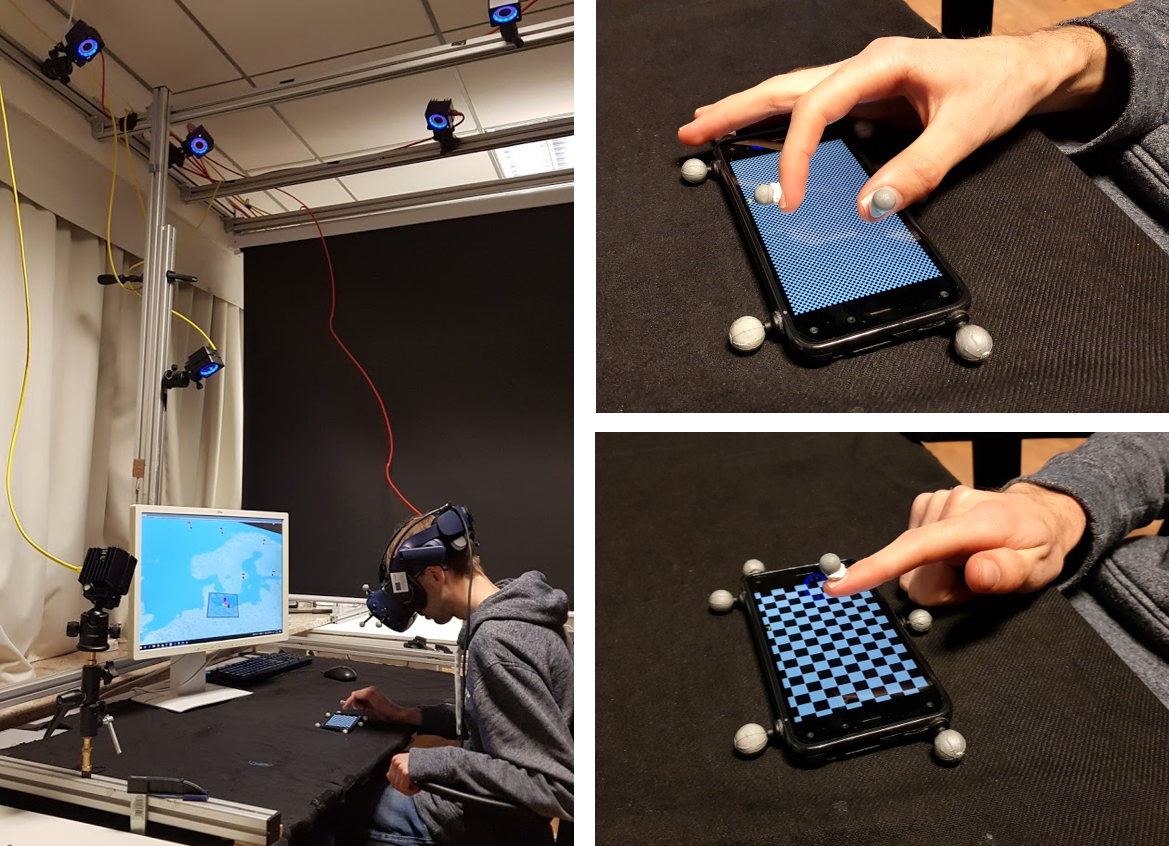}
	\caption{Left: Study location with OptiTrack Prime 13 and a user wearing the HTC VIVE Pro and interacting with the smartphone. The monitor shows the user's view of the scene. Top right: user pinches to zoom in condition \textsc{baseline}. Bottom right: user zooms in condition \textsc{3D Navigation}.}
	\label{fig:apparatus}
\end{figure}

For capturing the touch input and sending it via WiFi to the main application on a PC, we used an Android application on an Amazon Fire Phone running Fire OS 4.6.3. The smartphone dimensions were 139.2~mm × 66.5~mm × 8.9~mm (width x height x depth), with a touchscreen of size 105~mm x 60~mm. For 3D Tracking, we used an OptiTrack Prime 13 system, consisting of 8 cameras, of which 3 were Prime 13w wide field-of-view cameras and Motive Version 1.10.3.
The main study application was a running on a PC with windows 10, an Intel Xeon E5 - 1650 CPU, a Nvidia GeForce GTX 1070 graphics card and 64 GB RAM. The VR Headset used was the HTC Vive Pro. Both applications, the one on the smartphone as much as the one running on the PC were created in Unity 2018.3.2f1. While the HTC Vive Pro and the fingertips of users could have been tracked with mobile sensing solutions \cite{schneider2020accuracy} the HMD and fingers were equipped with retro-reflective markers for increased accuracy, see Figure \ref{fig:apparatus}.

\subsection{Participants}

We recruited 20 participants from a university campus with diverse study backgrounds. All participants were familiar with touch sensitive screens.  From the 20 participants (7 female, 13 male, mean age 24.4 years, sd = 2.973, mean height 175.8 cm, sd = 9.463, 15 indicated prior Virtual Reality Experience, 5 of those only once, 3 participants rarely but more than once, 5 participants occasionally and 2 often. Three participants indicated they don't play video games, 1 participant only once, 4 rarely, 6 occasionally, 3 frequently and 3 very frequently. Five participants wore contact lenses or glasses with corrected to normal vision. Nineteen participants were right handed while 1 was left handed. Nineteen participants used their right hand to conduct the task and 1 used the left hand. For  zooming in the 2D \textsc{baseline} interface, 15 participants used their thumb together with their index finger and five participants used their index finger in combination with their middle finger.

\subsection{Results}
%Statistical significance tests for target acquisition time and subjective feedback was carried out using general linear model repeated measures analysis of variance with Holm-Bonferroni adjustments for multiple comparisons at an initial significance level $\alpha = 0.05$. We indicate effect sizes whenever feasible ($\eta^2_p$). \hl{Please note, that the use of parametric methods is warranted for the data at hand, also if the test assumptions (such as normality of data) is violated, without concern for getting incorrect conclusions \cite{norman2010likert}.} 

Statistical significance tests for log-transformed target acquisition time was carried out using general linear model repeated measures analysis of variance (RM-ANOVA) with Holm-Bonferroni adjustments for multiple comparisons at an initial significance level $\alpha = 0.05$. %\hl{Please note, that the use of parametric methods is warranted for the data at hand, also if the test assumptions (such as normality of data) is violated, without concern for getting incorrect conclusions \cite{norman2010likert}.} 
We indicate effect sizes whenever feasible ($\eta^2_p$). We had to exclude performance data (target acquisition time and errors) for one participant due to logging errors.

%For NASA TLX, or data that did not follow a normal distribution or could not be transformed to a normal distribution using the log-transform, we employed the Aligned Rank Transform before applying RM-ANOVA.% with Holm-Bonferroni adjustments for multiple comparisons using Wilcoxon signed-rank tests (such as number of errors, questionnaire ratings). 

For subjective feedback, or data that did not follow a normal distribution or could not be transformed to a normal distribution using the log-transform (errors), we employed the Aligned Rank Transform before applying RM-ANOVA.% with Holm-Bonferroni adjustments for multiple comparisons using Wilcoxon signed-rank tests (such as number of errors, questionnaire ratings). 

The results in the following sections can be summarized as follows: Participants acquired targets significantly faster with \textsc{3D Navigation} compared to \textsc{baseline}. \textsc{3D Navigation} also resulted in significantly higher SUS scores, significantly lower demand ratings (as indicated by Nasa TLX) and was preferred by more participants. No significant differences for the number of errors or for simulator sickness ratings were detected between conditions.

\subsubsection{Target Acquisition Time}

% \begin{figure*}[t!]
% 	\centering
% 	\includegraphics[width=2\columnwidth]{images/conditions.png}
% 	\caption{Left: View on the VR scene in the \textsc{3D Navigation} condition at a small scale. The active target is shown as blue dot, inactive targets as red dots. The pink sphere indicates the fingertip position, the white disc below the pivot point for zooming. The touch screen area of the smartphone is shown as semi-transparent blue rectangle with red border. Right: View on the VR scene in the \textsc{baseline} condition at 1:1 scale. The circular targets are complemented with 3D flag symbols to indicate that targets can be selected. The two fingertips used for zooming are indicated by yellow and pink spheres. The pivot point indicated by the white disc is visualized half-way between both fingertips. }
% 	\label{fig:conditions}
% \end{figure*}

The times to acquire individual targets averaged over all target distances are depicted in Figure \ref{fig:performance}, left (for individual distances, see Appendix).
An omnibus test revealed significance main effects of \textsc{interface} ($F_{1, 18} = 59.91$, $p < .001$, $\eta^2_p = 0.77$, $1-\beta = 1.0$), and (as expected) for \textsc{map size} ($F_{1, 18} = 199.05$, $p < .001$, $\eta^2_p = 0.92$, $1-\beta = 1.0$) and \textsc{target distance} ($F_{2, 36} = 45.83$, $p < .001$, $\eta^2_p = 0.72$, $1-\beta = 1.0$). 

As expected, significant interactions have been indicated between \textsc{Interface} and \textsc{Map Size} ($F_{1,36} = 29.61$, $p < 0.001$, $\eta^2_p = 0.622$, $1-\beta = 0.99$), 
\textsc{Map Size} and \textsc{Target Distance} ($F_{2,36} = 11.08$, $p > 0.001$, $\eta^2_p = 0.38$, $1-\beta = 0.99$), but not between \textsc{Interface} and \textsc{Target Distance} ($F_{2,36} = 3.06$, $p = 0.059$, $\eta^2_p = 0.15$, $1-\beta = 0.56$). We did not observe any asymmetrical effects \cite{poulton1966unwanted}.
 
%An omnibus test revealed significance main effects of \textsc{interface} ($F_{1, 18} = 26.28$, $p < .001$, $\eta^2_p = 0.59$, $1-\beta = 1.0$), and (as expected) for \textsc{map size} ($F_{1, 18} = 76.86$, $p < .001$, $\eta^2_p = 0.81$, $1-\beta = 1.0$) and \textsc{target distance} ($F_{2, 36} = 18.15$, $p < .001$, $\eta^2_p = 0.50$, $1-\beta = 1.0$). 

%As expected, significant interactions have been indicated between \textsc{Interface} and \textsc{Map Size} ($F_{1,18} = 19.36$, $p < 0.001$, $\eta^2_p = 0.52$, $1-\beta = 0.99$), \textsc{Interface} and \textsc{Target Distance} ($F_{1,18} = 6.39$, $p = 0.004$, $\eta^2_p = 0.26$, $1-\beta = 0.88$), but not between \textsc{Map Size} and \textsc{Target Distance} ($F_{1,18} = 1.14$, $p = 0.33$, $\eta^2_p = 0.06$, $1-\beta = 0.23$), and not between \textsc{Interface}, \textsc{Map Size} and \textsc{Target Distance} ($F_{2,36} = 9.09$, $p = 0.81$, $\eta^2_p = 0.12$, $1-\beta = 0.08$). We did not observe any asymmetrical effects \cite{poulton1966unwanted}.

Holm-Bonferroni adjusted post-hoc testing revealed that there were significant differences between each level for target distance ($p < 0.001$, as expected), map size ($p < 0.001$, as expected) and between \textsc{baseline} (mean target acquisition time over all map sizes and target distances: 59.62 seconds, sd = 23.95) and \textsc{3D Navigation} (mean target acquisition time over all map sizes and target distances: 33.52 seconds, sd = 8.10) ($p < 0.001$).

Also, Holm-Bonferroni adjusted pairwise t-tests between \textsc{baseline} and \textsc{3D Navigation} for each combination of \textsc{map size} and \textsc{target distance} indicated that \textsc{3D Navigation} resulted in significantly faster target acquisition times compared to \textsc{baseline}.

Similar results were obtained when excluding one participant whose target acquistion times were substantially larger than the times of the other participants (depicted as outlier in Figure \ref{fig:performance}, left). These results are omitted for brevity. 

In other words, participants acquired targets significantly faster with \textsc{3D Navigation}  compared to \textsc{baseline}.
 
%Holm-Bonferroni adjusted post-hoc testing revealed significant differences between baseline and all randomization layouts (which was to be expected) (adjusted p-values $< 0.001$),

\subsubsection{Errors}
We looked at two different error types. First, we looked at the number of wrongly selected targets. No target selection error was made in any condition. Second, we counted the occasions when a target was not hit at first touch down (i.e. the finger was dwelling outside of the target circle for more than a second), see Figure \ref{fig:performance}, right. No significant main effects or interactions were indicated.%An omnibus test revealed no significant main or interaction effects.  %A Friedman test revealed no statistically significant difference between conditions ($\chi^2(11) = 11.37$, $p = 0.41$).

In other words, no significant differences between \textsc{3D Navigation} and \textsc{baseline} were detected.

\subsubsection{Workload}
The descriptive statistics for workload (as measured by the unweighted NASA TLX \cite{hart1988development}) are depicted in Table \ref{tab:tlx}.

 Significant main effects for \textsc{interface} were indicated for mental demand ($F_{1, 19} = 4.71$, $p = 0.043$, $\eta^2_p = 0.20$). physical demand ($F_{1, 19} = 6.42$, $p = 0.02$, $\eta^2_p = 0.25$), temporal demand ($F_{1, 19} = 22.82$, $p < 0.01$, $\eta^2_p = 0.55$), performance ($F_{1, 19} = 39.422$, $p < 0.01$, $\eta^2_p = 0.67$), effort ($F_{1, 19} = 18.30$, $p < 0.01$, $\eta^2_p = 0.49$), frustration ($F_{1, 19} = 66.18$, $p < 0.01$, $\eta^2_p = 0.78$) and overall demand ($F_{1, 19} = 39.95$, $p < 0.01$, $\eta^2_p = 0.68$). For \textsc{Map Size} a significant main effect was indicated for overall demand ($F_{1, 19} = 5.44$, $p = 0.031$, $\eta^2_p = 0.22$). No further main effects were indicated.
 
 Significant interactions were indicated for performance ($F_{1, 19} = 22.79$, $p < 0.01$, $\eta^2_p = 0.55$), effort ($F_{1, 19} = 7.78$, $p = 0.01$, $\eta^2_p = 0.29$), frustration ($F_{1, 19} = 4.89$, $p = 0.039$, $\eta^2_p = 0.20$) and overall demand ($F_{1, 19} = 11.77$, $p < 0.01$, $\eta^2_p = 0.38$).

 %Friedman tests revealed statistically significant difference between the conditions for all TLX dimensions (mental demand: $\chi^2(20) = 10.91$, $p = 0.012$, physical demand: $\chi^2(20) = 12.88$, $p = 0.005$, temporal demand: $\chi^2(20) = 11.66$, $p = 0.009$, performance: $\chi^2(20) = 32.95$, $p < 0.001$, effort: $\chi^2(20) = 13.42$, $p = 0.004$, frustration: $\chi^2(20) = 34.69$, $p < 0.001$, overall demand: $\chi^2(20) = 28.09$, $p < 0.001$).  Holm-Bonferroni adjusted post-hoc testing with Wilcoxon signed-rank tests revealed significant differences between \textsc{baseline} and \textsc{3D Navigation} for both map sizes ($p < 0.005$) as well as for overall demand for both map sizes ($p < 0.004$). For \textsc{large map}, significant differences between \textsc{baseline} and \textsc{3D Navigation} were indicated for temporal demand, performance and effort. No pairwise significant difference were indicated for mental demand or physical demand.
 
 In other words, \textsc{3D Navigation} led to a significantly lower workload compared to \textsc{baseline}.
 
 %In other words, \textsc{baseline} led to a significantly higher frustration and overall demand compared to \textsc{3D Navigation} for both \textsc{map size}s and to significantly higher temporal demand, performance and effort on the \textsc{large map}. 
 
 %temporal demand between \textsc{baseline} and \textsc{3D Navigation} for \textsc{large map} ($p < 0.001$)
 
  %performance between \textsc{baseline} and \textsc{3D Navigation} for \textsc{large map} ($p < 0.001$)
  
   %effort between \textsc{baseline} and \textsc{3D Navigation} for \textsc{large map} ($p = 0.002$)

 %frustration between \textsc{baseline} and \textsc{3D Navigation} for both map sizes ($p < 0.005$)
 
 %overall demand between \textsc{baseline} and \textsc{3D Navigation} for both map sizes ($p < 0.004$)

\begin{table}
	\caption{Mean and standard deviation (in parentheses) for the NASA TLX dimensions. 2D: \textsc{baseline}, 3D: \textsc{3D Navigation}, SM: \textsc{small map}, LM: \textsc{large map}.  Bold headings indicate dimensions with significant differences.}% The Bonferroni-corrected alpha level is 0.0167. Degrees of freedom = 22. Bold rows indicate significant differences.}
	\begin{center}
		%\hspace{-0.9cm}
		\begin{tabular}{ |c|c|c|c|c| }
			\hline
			%\multicolumn{4}{|c|}{ \textsc{Horizontal} Orientation } \\
		
			& 2D-SM & 3D-SM & 2D-LM & 3D-LM \\
			\hline
			    Mental Demand & 38.50 & 28.25 & 41.75  & 30.75  \\
			               & (26.01) & (16.33) & (23.64) & (19.89) \\
			\hline
			      Physical Demand          & 26.00 & 19.50 & 32.75  & 18.5  \\
			               & (21.56) & (17.24) & (25.73) & (16.47) \\
            \hline
			      \textbf{Temporal Demand}  & 41.00 & 31.50 & 55.50  & 30.25  \\
			                        & (25.00) & (19.54) & (30.78) & (21.30) \\
			 \hline
			      \textbf{Performance}       & 37.75 & 32.25 & 58.25  & 25.25  \\
			                        & (23.98) & (16.74) & (23.64) & (19.50) \\
            \hline
			      \textbf{Effort}              & 35.75 & 25.25 & 49.5 & 24.75  \\
			                        & (25.41) & (18.03) & (23.84) & (21.30) \\
			 \hline
			      \textbf{Frustration}       & 31.00 & 17.75 & 40.00  & 15.25  \\
			                        & (20.81) & (12.41) & (19.53) & (12.08) \\ 
            \hline
			      \textbf{Overall Demand}    & 35.00 & 25.75 & 46.29  & 24.13  \\
			                        & (16.96) & (10.88) & (14.80) & (12.38) \\ 
			 \hline
		\end{tabular}
	\end{center}
	%\caption{Table to test captions and labels}
	%\vspace{-0.5cm}
	\label{tab:tlx}
\end{table}

\subsubsection{Usability}

\begin{figure}[t!]
	\centering
	\includegraphics[width=\columnwidth]{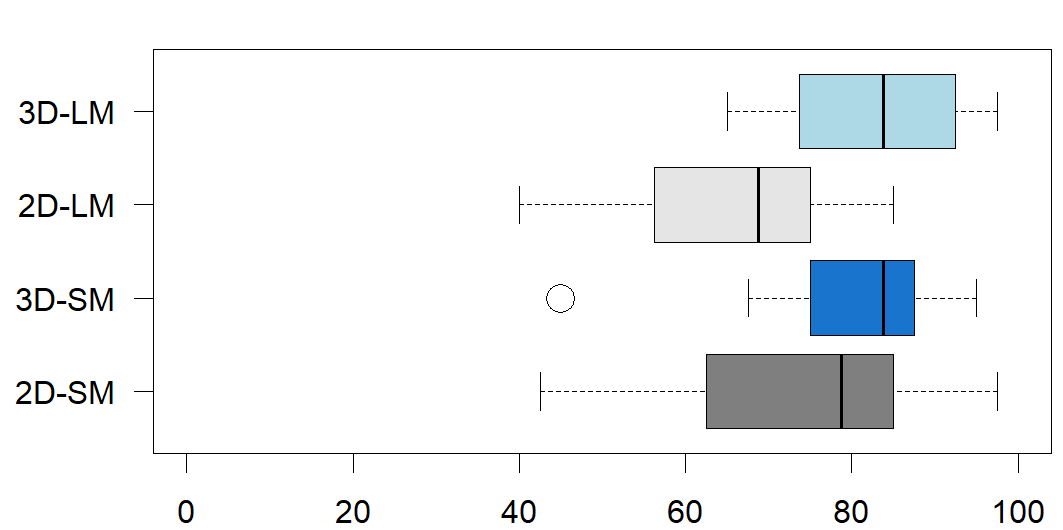}
	\caption{Ratings for the system usability scale SUS. 2D: \textsc{baseline}, 3D: \textsc{3D Navigation}, SM: \textsc{small map}, LM: \textsc{large map}.}
	\label{fig:sus}
\end{figure}

Results from the SUS questionnaire \cite{brooke1996sus} are depicted in Figure \ref{fig:sus}. All conditions but \textsc{baseline}-\textsc{Large Map} resulted above average SUS ratings $>$ 68 (with \textsc{baseline}-\textsc{Large Map} resulting in a mean score of 66).  

%A Friedman test revealed a statistically significant difference between the conditions ($\chi^2(20) = 26.76$, $p < 0.001$). Holm-Bonferroni adjusted post-hoc testing with Wilcoxon signed-rank tests revealed significant differences between \textsc{baseline} and \textsc{3D Navigation} for \textsc{Large Map} ($p < 0.001$) but not for \textsc{Small Map} ($p = 0.03$).

%In other words, \textsc{3D Navigation} resulted in significantly higher usability scores than \textsc{baseline} for the \textsc{Large Map} size.
 
A significant main effect for \textsc{interface} was indicated for the SUS score ($F_{1, 19} = 28.14$, $p < 0.001$, $\eta^2_p = 0.60$). A significant interaction between \textsc{Interface} and \textsc{Map Size} was also indicated ($F_{1, 19} = 6.09$, $p = 0.02$, $\eta^2_p = 0.24$).

In other words, \textsc{3D Navigation} resulted in significantly higher usability scores than \textsc{baseline}. 

\subsubsection{Simulator Sickness}

\begin{figure}[t!]
	\centering
	\includegraphics[width=\columnwidth]{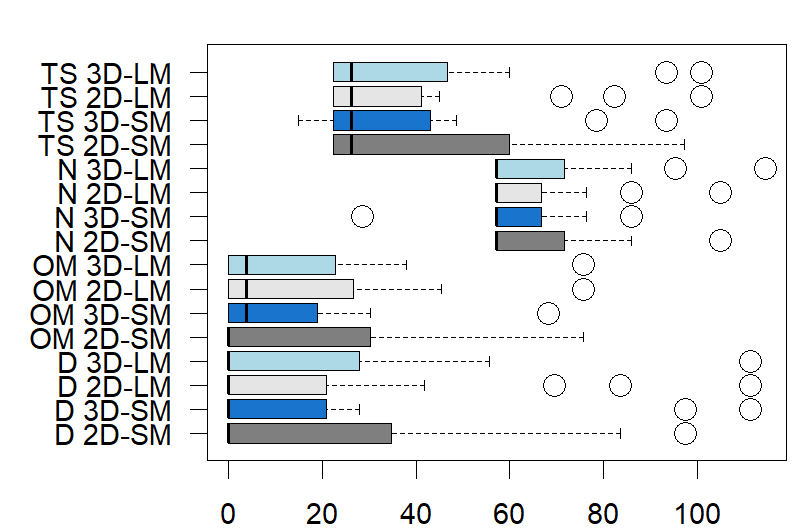}
	\caption{Ratings for the simulator sickness questionnaire SSQ. D: Disorientation, N: Nausea, O: Oculo-Motor, TS: Total Severity, 2D: \textsc{baseline}, 3D: \textsc{3D Navigation}, SM: \textsc{small map}, LM: \textsc{large map}.}
	\label{fig:ssq}
\end{figure}

Results from the simulator sickness questionnaire SSQ \cite{kennedy1993simulator} are depicted in Figure \ref{fig:ssq}. Omnibus tests revealed no statistically significant main effects or interactions for any SSQ dimension.% (Nausea: $\chi^2(20) = 1.27$, $p = 0.73$, Oculo-Motor: $\chi^2(20) = 2.66$, $p = 0.45$, Total Severity: $\chi^2(20) = 3.34$, $p = 0.34$, Disorientation: $\chi^2(20) = 1.77$, $p = 0.62$).

In other words, no significant difference for simulator sickness could be indicated between \textsc{3D Navigation} and \textsc{baseline}.

\subsubsection{Preferences and Open Comments}

When asked to rank the interaction techniques, all 20 participants preferred \textsc{3D Navigation} for \textsc{large map}. For \textsc{small map}, 15 participants preferred \textsc{3D Navigation} and 5 preferred \textsc{baseline}.

As benefits of \textsc{3D Navigation}, 8 participants mentioned it to be faster with one saying “I have the feeling that it is faster" and two mentioning it was also fast to learn. Three participants mentioned that they felt it enabled more precise navigation. Five participants it was easier to learn and "more fluid" with one mentioning “With 2D [\textsc{baseline}] I need more steps to zoom in or out. 3D [\textsc{3D Navigation}] just feels better” and four mentioning that using only a single finger for navigation felt more intuitive than the 2 finger pinch gesture.

Regarding drawbacks of the \textsc{3D Navigation} technique one participant mentioned that the “fast zooming gave me slight dizziness on the large map, but all in all it’s better, because it is faster”. Another participant mentioned  “I was losing the feeling on zoom range on single colored map parts” and 2 participants mentioned that they accidentally touched the screen, which led to an interruption of the zoom process.

Regarding benefits of  \textsc{baseline}, one participant mentioned that this technique helped better in the final stage of the target acquisition phase by stating “For me it was more precise because the frame is stable in the last step of target acquisition”. Two participants mentioned that they liked that a hovering finger does not invoke zooming. Three participants mentioned that they like the technique because they were used to it and three mentioned that they felt the technique to be fast enough on the small map. In contrast, two participants also mentioned that the technique was too slow for navigating on the large map as well as to imprecise. Another participant mentioned “my fingertips start to hurt if I touch the displays or things [in \textsc{baseline}] for too long, 3D on the other hand feels nice”.

We further noticed differences regarding the amount of zooming used by the participants in the different conditions. Overall, the participants zoomed less on \textsc{small map} for both interfaces. On a normalized scale, where 0 equals the minimum zoom scale (i.e. the whole map is visible within the smartphone boundaries) and 1 equals the maximum scale, the average scale across all participants was 0.425 for 2D Navigation and 0.357 for 3D Navigation on \textsc{small map}, compared to 0.065 and 0.064 for 2D Navigation and 3D Navigation on \textsc{large map} (please remember that participants started the task at the maximum scale 1 and had to be at that scale in order to select a target). In 70 instances for \textsc{baseline} respectively 30 instances for \textsc{3D Navigation} on \textsc{small map}, the participants found and clicked the highlighted target without zooming out at all. However, this was never the case in any of both conditions on \textsc{large map}.  Those differences can be attributed to the fact that on \textsc{small map}, a larger proportion of the map was already in the participants field of view when zoomed in. 

%0.575 for 2D Navigation and 0.643 for 3D Navigation on the small map size, compared to 0.935 and 0.936 for 2D Navigation and 3D Navigation on the large map size. In 70 respectively 30 cases for the 2D respectively the 3D Navigation on the small map size, the participants found and clicked the highlighted target without zooming out at all. However, this was never the case in any of both conditions on the large map. Those differences can be attributed to the fact that on the smaller map size, a larger proportion of the map was already in the participants field of view when zoomed in. 

%Regarding the use of the VR HMD, we observed two strategies. \hl{XXX} participants actively used the HMD to look around the virtual map, with \hl{XXX} only doing it in the \textsc{3D Navigation} condition. On the other hand \hl{XXX} participants, did not make extensive use of looking around the map with the HMD with one stating "I don’t use the capabilities of the VR-Headset, I don’t look around, It’s easier to zoom out look for the target and zoom in".

%\hl{TIM: can you find out the number of participants who used the HMD to look around and did not zoom out fully vs. the number of participants who zoomed out fully? Ideally separate the numbers for 2d vs 3d navigation}.

\section{Discussion and Limitations}

Our study indicated that the proposed relative, rate-controlled technique outperformed the \textsc{baseline} technique in terms of task completion time on both \textsc{map size}s. No differences in terms of errors were detected. While we assumed that potentially, differences between the conditions could appear for simulator sickness due to the faster scale change in \textsc{3D Navigation}, this assumption was not confirmed by the SSQ ratings.

The indications from workload and usability ratings were more nuanced. While for workload, \textsc{3D Navigation} resulted in significantly lower frustration and overall demand for both \textsc{map size}s, for the TLX dimensions temporal effort, performance and effort significant differences were solely indicated for the \textsc{large map}. No significant differences were indicated for mental or physical demand. One potential explanation for this might be that while holding the finger in mid-air (even with support of the palm lying on a resting surface) might induce fatigue, so does the repetitive use of pinch-to-zoom gestures.

Also, regarding usability ratings, \textsc{3D Navigation} was solely rated significantly higher for the \textsc{large map}. These quantitative results are also echoed by the qualitative feedback. While for \textsc{large map} all participants preferred \textsc{3D Navigation}, for \textsc{small map}, five out 20 preferred the \textsc{baseline} condition. This is also reflected by one participant stating "2D [baseline] on the small map is manageable but on the big map it is too much trouble".

With these findings in mind, we can still suggest the proposed relative, rate-control navigation technique can serve as viable option for navigating multiscale planar information spaces on and above touchscreens. 

As a limitation we can see, that in the experiment, the smartphone was lying on a surface, which also functioned as a resting surface for both interaction techniques. While this represents well the intended target usage in physical constrained spaces such as an airplane seat with a tray, other use cases, such as free hand usage or usage while walking could result in other findings, specifically regarding expected fatigue of the techniques. Further, in such scenarios, approaches that utilize the touchscreen pose for spatially navigating an information space (such as the work by Spindler et al. \cite{spindler2014pinch}) could be used and potentially outperform the proposed approach. Also, the results might change if other form factors such as tablets will be employed. Further, while we took care to select a representative range of information space sizes, the results might change when investigating further scales. Finally, one could argue that given spatial finger tracking on modern VR headsets, there is no need for a supporting touchscreen at all. While this is true for spatial interaction, and, eventually for sliding motions on physical surfaces such as tables, selecting items through nuanced touches on the surface without touchscreens (or other contact-based sensors) will remain a challenge for the foreseeable future. Not using a touchscreen also strips away further potential interaction possibilities (which we did not investigate in this work) that could arise due to the tangible nature of smartphones and tablets \cite{surale2019tabletinvr}.

\section{Conclusion}

The combination of physical touchscreen devices such as smartphones and tablets with Virtual Reality headsets enables unique interaction possibilities such as the exploration of large information spaces.

In this context, a controlled laboratory study with 20 participants indicated that a relative, rate-controlled multiscale navigation technique resulted in significantly better performance and user preference compared to pinch-to-zoom and drag-to-pan when navigating planar information spaces.

In future work, we plan to investigate the technique in further usage scenarios (such when standing or walking) and to explore the rich interaction space that opens up when combining traditional compute devices such as smartphones, tablets or notebooks with VR headsets capable of spatial tracking of the HMD, traditional input devices such as smartphones and the user's hands.
\balance
\bibliographystyle{abbrv-doi}

\bibliography{multiscale}
\end{document}